\documentclass[a4paper,
              ]{jacow}
\makeatletter%
	\ifboolexpr{bool{xetex}}
	 {\renewcommand{\Gin@extensions}{.pdf,%
	                    .png,.jpg,.bmp,.pict,.tif,.psd,.mac,.sga,.tga,.gif,%
	                    .eps,.ps,%
	                    }}{}
\makeatother

\ifboolexpr{bool{xetex} or bool{luatex}} % test for XeTeX/LuaTeX
 {}                                      % input encoding is utf8 by default
 {\usepackage[utf8]{inputenc}}           % switch to utf8

\usepackage[USenglish]{babel}			 

\usepackage[final]{pdfpages}
\usepackage{multirow}
\usepackage{booktabs}
\usepackage{ragged2e}
\usepackage{hyperref,graphicx,adjustbox}

\ifboolexpr{bool{jacowbiblatex}}%
 {%
  \addbibresource{jacow-test.bib}
  \addbibresource{biblatex-examples.bib}
 }{}
\begin{document}

\title{Proton beam dynamics in bare IOTA with intense space-charge}
\author{N. Banerjee\thanks{nilanjan@fnal.gov}, A. Romanov, M. Wallbank, Fermilab, Batavia, Illinois, USA}
		
\maketitle
\begin{abstract}
    We are commissioning a 2.5~MeV proton beam for the Integrable Optics Test Accelerator at Fermilab, allowing experiments in the strong space-charge regime with incoherent betatron tune shifts nearing 0.5. Accurate modelling of space-charge dynamics is vital for understanding planned experiments. We compare anticipated emittance growth and beam loss in the bare IOTA configuration using transverse space-charge models in Xsuite, PyORBIT, and MAD-X simulation codes. Our findings reveal agreement within a factor of 2 in core phase-space density predictions up to 100~synchrotron periods at moderate beam currents, while tail distributions and beam loss show significant differences.
\end{abstract}

\section{Introduction}
The Integrable Optics Test Accelerator (IOTA)\cite{Antipov2016} at Fermilab serves as a versatile platform for beam physics research, addressing core challenges\cite{Blazey2023} outlined in the Accelerator and Beam Physics Roadmap workshop. Our experiments focus on enhancing intensity and phase-space density within hadron synchrotrons and storage rings, employing 2.5~MeV protons\cite{Edstrom2023} under intense space-charge conditions. Our research objectives include validating Non-linear Integrable Optics\cite{Danilov2010, Wieland2023}, exploring the interplay between space-charge effects and coherent instabilities\cite{Ainsworth2021}, demonstrating bunch compression, and evaluating space-charge compensation, non-linear optics, and cooling using an electron lens\cite{Stancari2021}. Accurate space-charge modeling is crucial as we approach incoherent betatron tune shifts of up to 0.5. A previous effort of comprehensive benchmarks with several space-charge codes\cite{Schmidt2016} have used the SIS-18 heavy ion synchrotron in GSI operating at 11.4~MeV/u at a maximum tune-shift of 0.1. In contrast, this contribution compares projected beam properties of 2.5~MeV protons within IOTA in a strong space-charge regime, using three tracking codes—Xsuite\cite{iadarola2023}, PyORBIT\cite{Shishlo2015}, and MAD-X\cite{DeMaria2023}.\\

We utilize a \textit{bare} lattice configuration without non-linear elements, and with sextupoles deactivated for our benchmarking studies. Table~\ref{tab:params} presents the pertinent beam and lattice parameters. Our setup boasts unique characteristics including a low beam velocity of $c\beta_0 \approx 0.073 c$, which amplifies space-charge effects while rendering vacuum chamber impedance negligible. Additionally, we operate with a relatively short synchrotron period of $1/Q_s = 100$~turns, resulting in rapid betatron tune modulations of individual particles, and with a bunch length longer than the dipoles, resulting in banana-shaped beams in bend sections. We conduct assessments of emittance growth, beam loss, and transverse distributions after 100 synchrotron periods at varying intensities, spanning from negligible tune shift to crossing the closest integer resonance.\\

\begin{table}
    \caption{Parameters for benchmarking simulations with 2.5~MeV proton beams in IOTA.}
    \label{tab:params}
    \centering
    \begin{tabular}{lcl}
        \toprule
        \textbf{Parameter} & \textbf{Nominal} & \textbf{Unit} \\
        \midrule
        Kinetic energy ($K$) & 2.5 & MeV \\
        Geometric RMS emittances ($\epsilon_x, \epsilon_y$) & 4.3, 3 & $\mu$m \\
        RMS momentum spread ($\sigma_{\delta}$) & 2.1 & $10^{-3}$\\
        RMS bunch Length ($\sigma_z$) & 1.24 & m \\
        \midrule
        Circumference ($C$) & 40 & m\\
        RF harmonic ($h$) & 4 \\
        Working point ($Q_x, Q_y, Q_s$) & 5.3, 5.3, 0.01 \\
        Acceptances ($\epsilon_{mx}, \epsilon_{my}$) & 99, 120 & $\mu$m \\
        \bottomrule
    \end{tabular}
\end{table}

In the subsequent section, we detail the space-charge models and setup employed with each tracking code. Following this, we analyze the simulation results and explore their implications for proton operations. Finally, we summarize our findings and outline the next steps in our research.\\

\section{Space-charge codes and setup}
The presence of space-charge fields induces periodic modulation of betatron tunes\cite{Franchetti2003} in individual particles undergoing synchrotron motion within a bunched beam. In the absence of significant non-linearities and magnet errors, the periodic crossing of integer resonances becomes the primary driver of emittance growth, halo formation, and beam loss. This paper specifically benchmarks this scenario within the \textit{bare} lattice configuration of IOTA, devoid of magnetic field errors, and with a nominal aperture of 25~mm. Longitudinal space-charge forces are disregarded due to the substantially larger longitudinal size of the bunch compared to the transverse size. Therefore, we utilize codes that incorporate transverse space-charge kicks, considering the local charge density as a function of longitudinal position within the bunch.\\

We compare results from three distinct space-charge models: the frozen model in Xsuite, the quasi-frozen model in MAD-X, and the 2.5D Particle-in-Cell (PIC) model in PyORBIT. Both the frozen and quasi-frozen models assume that the beam is spatially well-described by a 3D Gaussian distribution, employing a semi-analytical approach to calculate transverse kicks. The quasi-frozen model updates the rms sizes of the bunch as it propagates in the lattice, thereby updating the space-charge fields, while the frozen model assumes constant fields. In contrast, the PIC model solves Poisson's equation in the rest-frame of the bunch to calculate self-fields, which are then used to update the transverse momenta of macro-particles. Consequently, the PIC model can represent arbitrary particle distributions but requires significantly greater computational effort. We employ $10^4$ macro-particles for Xsuite and MAD-X simulations, and $10^5$ macro-particles with a PIC grid resolution of $128^3$ for PyORBIT simulations.\\

\begin{figure}
    \centering
    \includegraphics[width=\columnwidth]{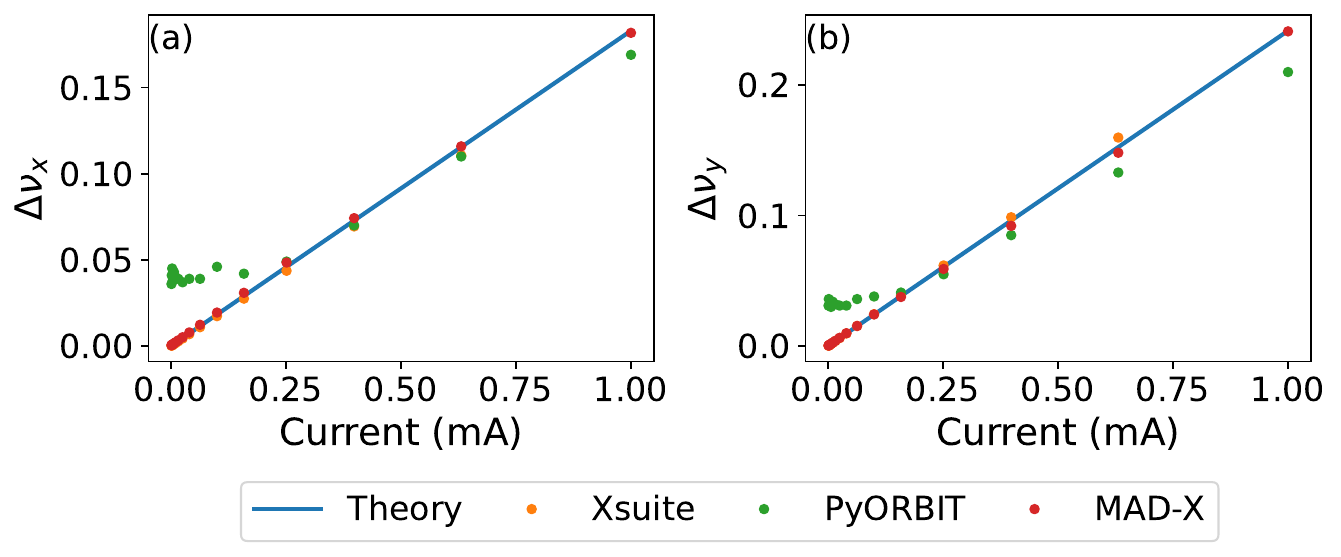}
    \caption{Transverse incoherent tune shifts due to space-charge as a function of beam current in IOTA. Panels (a) and (b) show the horizontal and vertical tune shifts respectively. The dots represent results from three simulation codes while the blue line denotes the expected value given by Eq.~(\ref{eq:dnu}).}
    \label{fig:tuneshift}
\end{figure}

The initial probability density function describing the beam is Gaussian in all directions, truncated to $3\sigma$ and $2.5\sigma$ for the transverse and longitudinal planes respectively, and matched with the linear optics of the lattice. For Xsuite and PyORBIT simulations, the Twiss parameters of the beam distribution match those of the bare lattice without considering space-charge tune shifts. However, MAD-X employs a self-consistent linear space-charge matching algorithm to determine modified linear lattice functions for the beam core, perturbed due to linear defocusing from space-charge, and adjusts the initial beam distribution accordingly. We validate our simulation setup by estimating transverse incoherent tune-shifts ($\Delta \nu_{\mathrm{SC},x,y}$) due to space-charge and comparing them with expected values given by\cite{Litvinenko2015}
\begin{equation}
    \label{eq:dnu}
    \begin{split}
        \Delta \nu_{\mathrm{SC},x,y} &=\frac{Nr_p}{(2\pi)^{3/2}\beta_0^2\gamma_0^3 h\sigma_z} \times \\
        &\int_0^C \frac{\beta_{x,y}(s)}{\{\sigma_x(s)+\sigma_y(s)\}\sigma_{x,y}(s)}\,\mathrm{d}s\,,
    \end{split}
\end{equation}
where $N=IC/(ec\beta_0)$ is the total number of protons in the ring, dependent on the average beam current $I$. Transverse beam sizes are given by $\sigma_{x,y}(s) = \sqrt{\beta_{x,y}(s)\epsilon_{x,y}+\sigma_\delta^2D_{x,y}^2(s)}$, where $\beta_{x,y}$ and $D_{x,y}$ represent transverse betatron amplitudes and dispersion respectively. Tune shifts in Xsuite and MAD-X simulations are estimated by introducing an extra macro-particle displaced from the center of the bunch by a nanometer, in addition to the matched bunch distribution, and measuring the tune of this particle. Conversely, individual particle amplitudes in PIC simulations tend to be chaotic with respect to initial conditions; therefore, we measure the tunes of all particles and find the particle with the smallest tune. Figure~\ref{fig:tuneshift} illustrates that tune-shifts estimated from Xsuite and MAD-X match within a percent level, while agreement with PyORBIT is within 10\% at high intensities, with chromatic tune shift dominating at low intensities. Utilizing this setup, we simulate proton beams in IOTA at four different currents, $I=0.001,\, 0.5,\, 1.0,\,\&\, 2.0$~mA, up to 100~synchrotron periods, corresponding to $T=100/Q_s = 10^4$ turns.\\

\section{Comparison of phase-space evolution}
The evolution of root mean square emittance and beam loss are crucial predictions of tracking codes, often validated through experiments. We use a definition of root mean square horizontal emittance $\epsilon_{\mathrm{eff},x}$, compensated for dispersive correlations, as used in PyORBIT:
\begin{subequations}
    \begin{align}
        \label{eq:rmsemittance}
        \epsilon_{\mathrm{eff},x} &\equiv \sqrt{\langle x^2 \rangle_\mathrm{eff} \langle x'^2 \rangle_\mathrm{eff} - \langle xx' \rangle_\mathrm{eff}^2}\,,\\
        \langle x^2 \rangle_\mathrm{eff} &\equiv \langle x^2 \rangle - \frac{\langle x\delta \rangle^2}{\langle \delta^2 \rangle}\,, \\
        \langle x'^2 \rangle_\mathrm{eff} &\equiv \langle x'^2 \rangle - \frac{\langle x'\delta \rangle^2}{\langle \delta^2 \rangle}\,, \\
        \langle xx' \rangle_\mathrm{eff} &\equiv \langle xx' \rangle - \frac{\langle x\delta \rangle \langle x'\delta \rangle}{\langle \delta^2 \rangle}\,,
    \end{align}
\end{subequations}
where $\langle\rangle$ denotes mean values, $x$ and $x'$ are horizontal phase-space coordinates, and $\delta$ represents the relative momentum deviation of particles in the beam. Similarly, we define $\epsilon_{\mathrm{eff},y}$	
to characterize the vertical phase-space. While root mean square quantities capture both the core and tails of the beam distribution, we introduce another metric $\epsilon^*_{x,y}$ to quantify the inverse phase-space density of the circulating beam at its core:
\begin{equation}
    \label{eq:epsilonstar}
     \epsilon^*_{x,y} = -\frac{f_{x,y}(0)}{2} \bigg( \frac{\mathrm{d}f_{x,y}}{\mathrm{d}J_{x,y}} \bigg\vert_{J_{x,y}=0} \bigg)^{-1}\,,
\end{equation}
where $f_{x,y}(J_{x,y})$ represents the transverse phase-space distribution of the bunch, and $J_x \equiv \beta_x^2 x'^2 +2\alpha_x xx'+\gamma_x x^2$ and $J_y \equiv \beta_y^2 y'^2 +2\alpha_y yy'+\gamma_y y^2$	are the horizontal and vertical actions of a single particle, assuming no transverse coupling. In the absence of dispersion, if the bunch has a pure Gaussian distribution, $\epsilon_{\mathrm{eff},x,y} = \epsilon^*_{x,y}$. We measure these metrics at the point of injection in the lattice where $\alpha_{x,y}=0$ while $D_x \neq 0$, and estimate the betatron amplitudes from the beam distribution as $\beta_{x} \approx \sqrt{\langle x^2 \rangle/\langle x'^2 \rangle}$ and $\beta_{y} \approx \sqrt{\langle y^2 \rangle/\langle y'^2 \rangle}$.\\

\begin{figure*}
    \centering
    \includegraphics[width=0.67\textwidth]{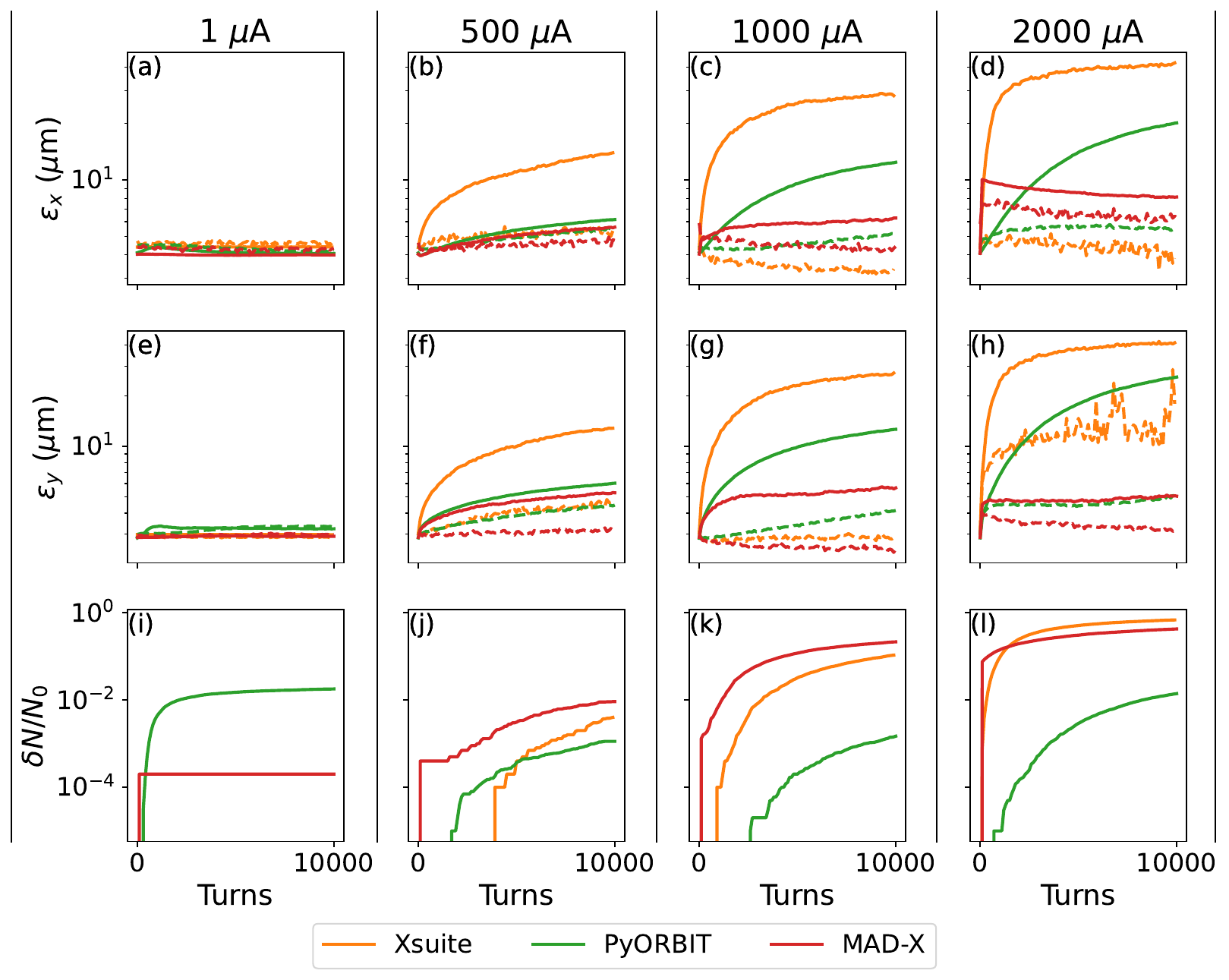}
    \caption{Comparison of transverse emittance and beam loss in IOTA as a function of turn number, between different simulation codes. Panels (a), (b), (c) and (d) indicate the evolution of horizontal emittance, while panels (e), (f), (g) and (h) display vertical emittance. Panels (i), (j), (k) and (l) present fractional beam loss. The 4 columns of the figure correspond to beam currents $1\,\mu$A, 0.5~mA, 1~mA and 2~mA in order. The solid lines denote rms emittance compensated for dispersion as defined in Eq.~(\ref{eq:rmsemittance}), while the dashed lines mark the inverse phase-space density defined in Eq.~(\ref{eq:epsilonstar}).}
    \label{fig:emittance}
\end{figure*}

We present the evolution of emittance metrics and beam loss over time, up to $10^4$ turns, at four different beam currents in Fig.~\ref{fig:emittance}. At $1\,\mu$A, the incoherent tune shift due to space-charge is $\sim 2\times10^{-4}$ in both transverse directions, resulting in linear dynamics where the phase-space distribution remains Gaussian and emittance growth is negligible, consistent across all three codes. At moderate beam currents corresponding to $\Delta\nu_{\mathrm{SC},x,y}<0.3$, substantial growth in transverse rms emittances ($\epsilon_{\mathrm{eff},x,y}$) accompanied by beam loss is observed in all codes, albeit with quantitative discrepancies. However, the inverse phase-space density ($\epsilon^*_{x,y}$) at the core remains within 30\% of its initial value in the results from all three tracking models, indicating consensus among different algorithms regarding the dynamics of the beam core at moderate currents. Once the incoherent tune shift exceeds 0.3, particles at the center of the beam cross the integer resonance, resulting in rapid beam blow-up, with all three simulations diverging on emittance growth rates and final emittance values. In summary, while the evolution of the beam core remains consistent among the three codes until particles in the core cross the integer resonance, there is no agreement beyond this point.\\

\section{Outlook}
An accurate model of space-charge is crucial for analyzing the beam dynamics of 2.5 MeV protons at the Integrable Optics Test Accelerator, especially in the intensity regime where the incoherent tune shift may reach 0.5 during experiments. Utilizing the frozen, quasi-frozen, and Particle-in-Cell (PIC) models of transverse space-charge in the tracking codes Xsuite, MAD-X, and PyORBIT respectively, we predict beam dynamics at four beam currents over 100 synchrotron periods. Our findings reveal substantial quantitative discrepancies in predicting root mean square emittances and beam loss at moderate and high currents among the three tracking models. However, consensus emerges in predicting the behavior of the beam core at moderate currents, as long as particles remain below the integer resonance. Building on these results, we aim to broaden our study by incorporating other PIC models, particularly those in IMPACT-X\cite{Mitchell2023}, Synergia\cite{Amundson2023}, and Xsuite. Additionally, we intend to conduct more detailed investigations into single-particle dynamics across various action values, along with comprehensive comparisons of beam distributions at different intensities. These efforts will provide a clearer understanding of the distinctions between tracking models and enhance our insights into space-charge dynamics.\\

\section*{Acknowledgement}
We would like to thank Chad Mitchell, Frank Schmidt, Ben Simons, Eric Stern and John Wieland for useful discussions and help with the code. This manuscript has been authored by Fermi Research Alliance, LLC under Contract No.~DE-AC02-07CH11359 with the U.S.\ Department of Energy, Office of Science, Office of High Energy Physics.


\begin{thebibliography}{99}
    \bibitem{Antipov2016}
    S.~Antipov \emph{et al.}, \textquotedblleft{IOTA (Integrable Optics Test Accelerator): Facility and Experimental Beam Physics Program}\textquotedblright, \emph{JINST} 12, T03002, 2017.
    \url{doi:10.1088/1748-0221/12/03/t03002}
    
    \bibitem{Blazey2023}
    J.~Blazey \emph{et al.}, \textquotedblleft{Accelerator and Beam Physics Roadmap}\textquotedblright, DOE Accelerator Beam Physics Roadmap Workshop, 2022.

    \bibitem{Edstrom2023}
    D.~R.~Edstrom \emph{et al.}, \textquotedblleft{Status of the IOTA Proton Injector}\textquotedblright, in \emph{Proc. HB'23}, Geneva, Switzerland, Oct. 2023, pp.~629--632. \url{doi:10.18429/JACoW-HB2023-FRA1I1}

    \bibitem{Danilov2010}
    V.~Danilov and S.~Nagaitsev, \textquotedblleft{Nonlinear accelerator lattices with one and two analytic invariants}\textquotedblright, \emph{Phys. Rev. ST Accel. Beams} vol.~13, 084002, 2010. \url{doi:10.1103/PhysRevSTAB.13.084002}

    \bibitem{Wieland2023}
    J.~Wieland \emph{et al.}, \textquotedblleft{Improved measurements of nonlinear integrable optics at IOTA}\textquotedblright, in \emph{Proc. IPAC’23}, Venice, Italy, May 2023, pp. 3230--3232. \url{doi:10.18429/JACoW-IPAC2023-WEPL052}

    \bibitem{Ainsworth2021}
    R.~Ainsworth \emph{et al.}, \textquotedblleft{A Dedicated Wake-Building Feedback System to Study Single Bunch Instabilities in the Presence of Strong Space Charge}\textquotedblright, in \emph{Proc. HB'21}, Batavia, IL, USA, Oct. 2021, pp. 135--139. \url{doi:10.18429/JACoW-HB2021-MOP22}

    \bibitem{Stancari2021}
    G.~Stancari \emph{et al.}, \textquotedblleft{Beam physics research with the IOTA electron lens}\textquotedblright, \emph{JINST} vol.~16, no.05, P05002, 2021. \url{doi:10.1088/1748-0221/16/05/p05002}

    \bibitem{Schmidt2016}
    F. Schmidt \emph{et al.}, \textquotedblleft{Code Bench-Marking for Long-Term Tracking and Adaptive Algorithms}\textquotedblright, in \emph{Proc. HB’16}, Malmö, Sweden, Jul. 2016, pp. 357--361. \url{doi:10.18429/JACoW-HB2016-WEAM1X01}

    \bibitem{iadarola2023}
    G.~Iadarola \emph{et al.}, \textquotedblleft{Xsuite: An Integrated Beam Physics Simulation Framework}\textquotedblright, in \emph{Proc. HB'23}, Geneva, Switzerland, Oct. 2023, pp.~73--80. \url{doi:10.18429/JACoW-HB2023-TUA2I1}

    \bibitem{Shishlo2015}
    A.~Shishlo \emph{et al.}, \textquotedblleft{The Particle Accelerator Simulation Code PyORBIT}\textquotedblright, \emph{Procedia Computer Science} vol.~51, pp.~1272--1281, 2015. \url{doi:10.1016/j.procs.2015.05.312}

    \bibitem{DeMaria2023}
    R.~De~Maria \emph{et al.}, \textquotedblleft{Status of MAD-X V5.09}\textquotedblright, in \emph{Proc. IPAC’23}, Venice, Italy, May 2023, pp. 3340--3343. \url{doi:10.18429/JACoW-IPAC2023-WEPL101}

    \bibitem{Franchetti2003}
    G.~Franchetti \emph{et al.}, \textquotedblleft{Space charge and octupole driven resonance trapping observed at the CERN Proton Synchrotron}\textquotedblright, \emph{Phys. Rev. ST Accel. Beams} vol.~6, 124201, 2003. \url{doi:10.1103/PhysRevSTAB.6.124201}

    \bibitem{Litvinenko2015}
    V.~N.~Litvinenko and G.~Wang, \textquotedblleft{Compensating tune spread induced by space charge in bunched beams}\textquotedblright, \emph{Phys. Rev. ST Accel. Beams} vol.~17, 114401, 2015. \url{doi:10.1103/PhysRevSTAB.17.114401}

    \bibitem{Mitchell2023}
    C.~E.~Mitchell \emph{et al.}, \textquotedblleft{ImpactX Modeling of Benchmark Tests for Space Charge Validation}\textquotedblright, in \emph{Proc. HB'23}, Geneva, Switzerland, Oct. 2023, pp.~583--586. \url{doi:10.18429/JACoW-HB2023-THBP44}

    \bibitem{Amundson2023}
    J.~Amundson \emph{et al.}, \textquotedblleft{SYNERGIA 2}\textquotedblright, \url{https://synergia.fnal.gov}.
    
\end{thebibliography}
\end{document}